\documentclass[english,prl,floatfix,superscriptaddress,twocolumn,showpacs,amsmath,amssymb,reprint]{revtex4-1}
\usepackage[T1]{fontenc}
\usepackage{color}
\usepackage{textcomp}
\usepackage{amsmath}
\usepackage{amssymb}
\usepackage{graphicx}
\usepackage{wasysym}
\usepackage{gensymb}

\makeatletter

\usepackage{bbold}
\usepackage{dsfont}
\PassOptionsToPackage{caption=false}{subfig} 
\usepackage{hyperref}
\hypersetup{
breaklinks=true,
colorlinks=true,
citecolor=blue,
linkcolor=blue,
filecolor=blue,
urlcolor=blue
}
\IfFileExists{lmodern.sty}{\usepackage{lmodern}}{}
\makeatother

\usepackage{babel}
\begin{document}
\title{Floquet engineering correlated materials with unpolarized light}
\author{V. L. Quito}
\email{vquito@iastate.edu}
\affiliation{Department of Physics and Astronomy, Iowa State University, Ames,
Iowa 50011, USA}
\author{R. Flint}
\affiliation{Department of Physics and Astronomy, Iowa State University, Ames,
Iowa 50011, USA}
\date{\today}
\begin{abstract}
Floquet engineering is a powerful tool that drives materials with periodic light. Traditionally, the light is monochromatic, with amplitude, frequency, and polarization varied.  We introduce Floquet engineering via unpolarized light built from quasi-monochromatic light, and show how it can modify strongly correlated systems, while preserving the original symmetries.  Different types of unpolarized light can realize different strongly correlated phases As an example, we treat insulating magnetic materials on a triangular lattice and show how unpolarized light can induce a Dirac spin liquid.

\end{abstract}
\maketitle

Floquet engineering provides a powerful method to access and control phases and phenomena absent or rare in equilibrium~\cite{Oka_review_2019,Mentink_NatPhys_2015,Mentink_2017,Ishihara_PRB_2018,Losada_PRB_2019,Millis_PRB_2019,Refael_PRB_2019,Mentink_2019_SciPost,Giustino_2021,Rudner2019}.  In its most common application, Floquet engineering consists of continuously driving a sample with monochromatic laser light, which has a fixed polarization. Unless the polarization axis is preserved by lattice symmetries, polarized light explicitly breaks either lattice (linear polarization), time-reversal (circular polarization) or both  symmetries.  This explicit symmetry breaking can be useful, as new couplings like chiral fields that induce spin chirality~\cite{sato_arxiv_2014,SatoOka_2016_PRL,Kitamura_PRB_2017,ClaassenNatComm2017} or anomalous Hall effects~\cite{Oka_2009_PRB,DemlerPRB2011,Lindner2011} can be generated; 
 or spatial anisotropies and dimensionalities can be tuned~\cite{Halperin_PRX_2017,Refael_PRL_2018,Yuan_Optics_2018,Ozawa_NatureRev_2019,Dutt_Science_2020}. 
While polarized light drives interesting physics \cite{Oka_review_2019,Wang2013Science,McIver2020}, some correlated phases are only accessible if all symmetries are preserved. For example, symmetric spin liquids require preserving lattice and time-reversal symmetries \cite{balents10}.  These phases are found in equilibrium models, but are confined to small regions of phase space theoretically, and are extremely rare experimentally. Floquet engineering could provide a new way to access spin liquids in materials, and to tune across their quantum critical points.

\begin{figure}
\includegraphics[width=1.0\columnwidth]{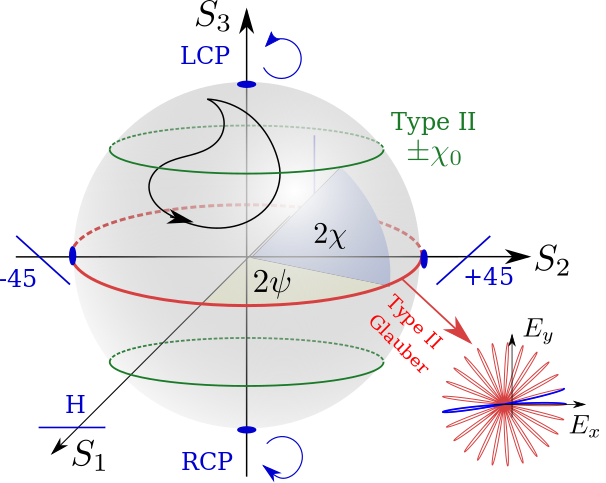}\caption{The Poincar\'{e} sphere captures all polarization profiles. The axes are the Stokes parameters, which give the degree of horizontal/vertical ($S_1$), $\pm  45^\circ$  ($S_2$) and circular ($S_3$) polarization. The monochromatic light traditionally used in Floquet engineering corresponds to a single point. Unpolarized light corresponds to \emph{paths} on the sphere with $\langle \vec{S}\rangle =0$. As we show, different paths can lead to distinct correlated phases. Any parallel of constant latitude, $\chi$ preserves rotational symmetry (e.g. - red curve). If parallels of both $\pm \chi$ are included (green curves), time-reversal is also preserved.  Inset: Parametric plot of the electric field of our simple example, where the polarization traverses the equator with period $T_p$. The thick blue line shows a single period $T$.
\label{fig:Poincar_sphere}}
\end{figure}

Unpolarized light preserves symmetries, but is not strictly monochromatic. Thus, it is not obvious that Floquet techniques apply, or what the effect is on correlated materials, although some promising work analyzed the effect of noise in Floquet engineered graphene~\cite{MukherjeePRB2018}.  This letter provides the general theory and applicability of unpolarized light in Floquet systems. Different kinds of unpolarized light sample polarizations differently, understood as different paths over the Poincar\'e sphere shown in Fig. \ref{fig:Poincar_sphere}. We prove that Floquet engineering with effectively unpolarized light is possible and introduce a simple model with two oppositely circularly polarized lasers whose frequencies are slightly detuned, resulting in unpolarized light whose polarization vector explores the equator of the Poincar\'e sphere.  We calculate the effect on magnetic exchange interactions in Mott insulators, and show that polarization averaging of the final result agrees with the exact result for sufficiently slow variation of the polarization vector. We then consider all types of unpolarized light and show how varied realistic choices can give significantly different exchange couplings while preserving the same symmetries. We treat the half-filled triangular Hubbard model in detail and show how to boost the ratio of $J_2/J_1$ and potentially access both the Dirac~\cite{hu15,zhu15,li15,iqbal16,saadatmand17,wietek17,gong17,gong_PRB_2019} spin liquid and the time-reversal symmetry breaking chiral spin liquid~\cite{messio13,hu16,wietek17,gong17}.  Finally, we discuss how Floquet engineering with unpolarized light may be reasonably implemented experimentally.

\begin{figure}
\includegraphics[width=.95\columnwidth]{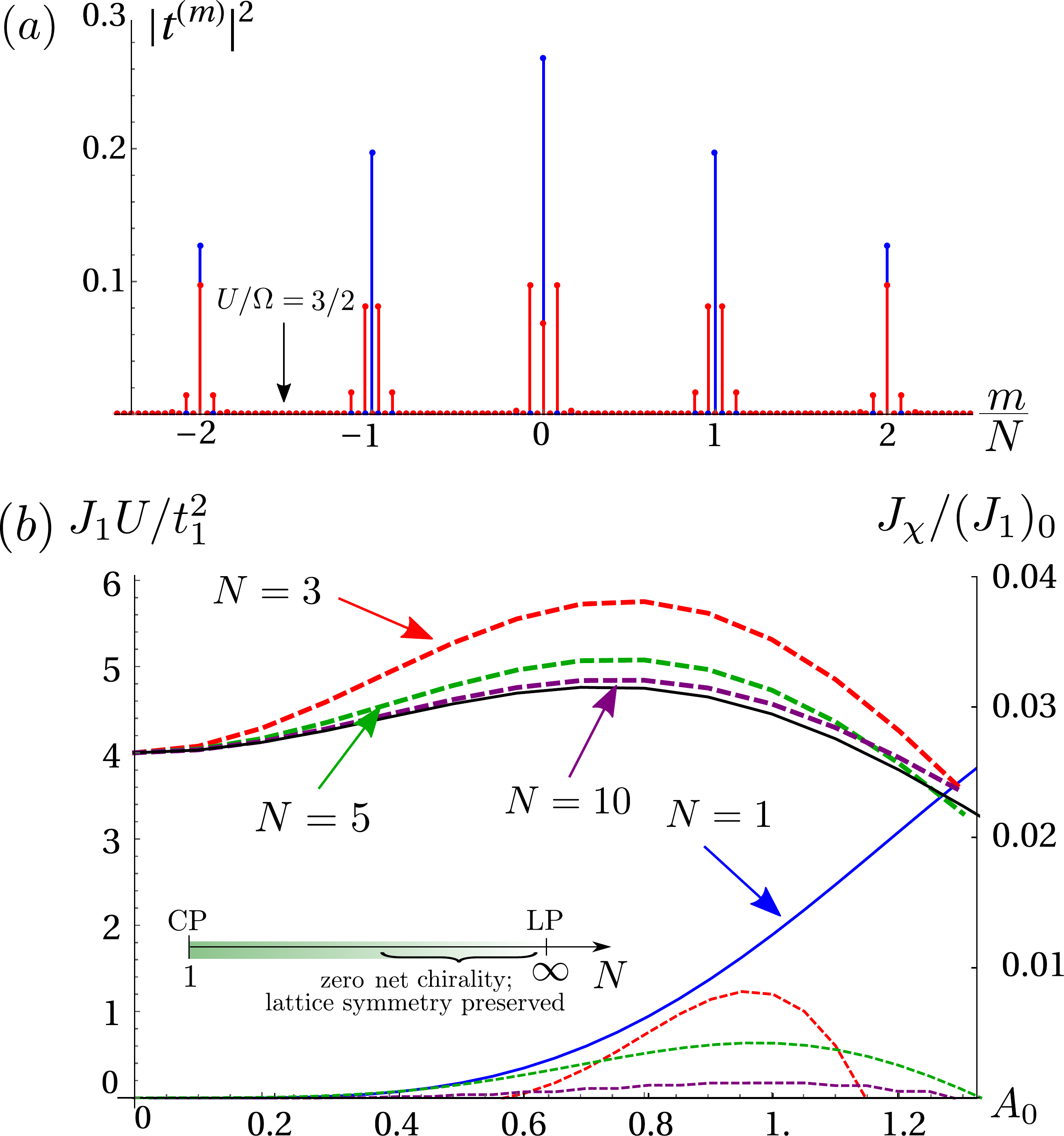}\caption{ (a) The hopping terms $\left|t^{\left(m\right)}\right|^{2}$ (red) as a function of $m$, for $N=25$ and $A_0=1.5$. The non-negligible values cluster around $m = N \tilde{m}$ with small satellite peaks.  When these clusters are well separated, as for sufficiently large $N$, the contribution of each cluster can be summed to give an approximate $\left|t^{\left(\tilde{m}\right)}\right|^{2}$ (blue). It is important to avoid resonances with non-negligible weight. The arrow indicates our chosen frequency with $m/N=-3/2$ for $N\Omega_{p}/U=\Omega/U=2/3$, where the  amplitude of $\left|t^{\left(m\right)}\right|^{2}$ is vanishingly small.  (b) $J_1$ as a function of fluence $A_0$.  The black line indicates the monochromatic average over LPs, which coincides with the $N \geq 25$ results. Also shown with dashed lines are chiral couplings on a triangular lattice, $J_{\chi}$, normalized by the bare $(J_{1})_{0}$. $N=1$ corresponds to the circularly polarized case. As $N$ increases, $J_{\chi}$ becomes vanishingly small.
\label{fig:contributions_t}}
\end{figure}

We examine magnetic exchange couplings in a single-band Floquet-Hubbard model, where electrons hop on a lattice in the presence of a time-dependent electric field, $\boldsymbol{E}=-\frac{\partial\boldsymbol{A}}{\partial t}$. There is a strong penalty for double occupancy, $U$:
\begin{equation}
\mathcal{H}_{0} =-t_{1}\!\sum_{i,\boldsymbol{\delta}_{i}}\mathrm{e}^{-i\boldsymbol{A}(t)\cdot\boldsymbol{\delta}_{i}}c_{i}^{\dagger}c_{i+\boldsymbol{\delta}_{i}}+U\!\sum_{i}n_{i\uparrow}n_{i\downarrow}-\mu\! \sum_{i} \! n_{i}.\label{eq:Hamiltonian}
\end{equation}
The chemical potential, $\mu$ is adjusted to ensure half-filling. We consider only nearest-neighbor links labeled by $\boldsymbol{\delta}_{i} = (\cos \phi_i, \sin \phi_i)$ and assume light propagation normal to the sample.  We take the vector potential to be time-periodic, with period $T = 2\pi/\Omega$, which allows the Fourier-transform to Floquet space with the \emph{discrete} set of frequencies \cite{Shirley_PR_1965,Sambe_PRA_1973,mahan2010book,AokiPRB2016}, $m \Omega$, $m\in \mathbb{Z}$.  In this space, the electrons now hop not just between sites, but between Floquet sectors labeled by $|m\rangle$ \cite{AokiPRB2016,Oka_review_2019}, 
\begin{align}
\mathcal{H} & =-\sum_{m,n}\sum_{i,\boldsymbol{\delta}_{i}}t_{i,i+\boldsymbol{\delta}_{i}}^{\left(n-m\right)}c_{i}^{\dagger}c_{i+\boldsymbol{\delta}_{i}}\left|m\right\rangle \left\langle n\right| \nonumber \\
 & +\sum_m\sum_i\left(U n_{i\uparrow}n_{i\downarrow}+m\Omega\right) |m\rangle\langle m|\label{eq:Hamiltonian-honeycomb-Floquet}
\end{align}
One key feature is that $\Omega$ may be tuned such that $m\Omega=-U$ for some integer $m$.  If so, pairs of doublons and holons will be excited across the Hubbard gap~\cite{Berges_PRL_2004,Abanin_mathem_2017,Mori_PRL_2016,RigolPRX2014,Lazarides_PRE_2014,Kuwahara_Annals_2016}, a resonance that destroys the Mott insulating state. However, for frequencies away from these resonances, which have width $\sim t_1$, the heating is minimal and an effective spin model treatment is justified \cite{ClaassenNatComm2017}.

Typically, the Floquet formalism treats a single frequency, $\Omega$, but we now extend it to quasi-monochromatic unpolarized light. We consider light combining two circularly polarized beams with slight frequency detuning, which causes the polarization vector to circle the equator, as shown in the inset of Fig.~\ref{fig:Poincar_sphere}, sampling all linear polarizations equally. We call the two frequencies $\Omega_{\pm} \equiv \Omega \pm \Omega_p$, and require that they be commensurate to ensure overall time-periodicity.  We assume that the period of the light, $T = 2\pi/\Omega$ is small compared to the period of the polarization, $T_p = 2\pi/\Omega_p$, and the commensurability is ensured by taking $T_p = N T$, where $N$ is an integer that is large for quasi-monochromatic light. The electric field is,
\begin{align}
    \boldsymbol{E}(t) & = E_0 \left(\begin{array}{c}\cos \Omega_p t\\\sin \Omega_p t\end{array}\right) \text{Re} [\mathrm{e}^{-i \Omega t}],\cr 
    & = \frac{E_0}{2} \text{Re}\left[ \left(\begin{array}{c}1\\i\end{array}\right)\mathrm{e}^{-i \Omega_+ t}+ \left(\begin{array}{c}1\\-i\end{array}\right)\mathrm{e}^{-i \Omega_- t} \right]. \label{eq:quasi-mon-protocol}
\end{align}

We can use perturbation theory to find the magnetic exchange couplings numerically for any integer $N$, and analytically for large, but finite $N$; exactly at $N = \infty$, the light is linearly polarized.  We first find the effective hoppings between sites and Floquet sectors, generically given by $t_{i,i+\boldsymbol{\delta}_{i}}^{\left(m\right)}  =t_{1}/(2\pi)\int_{0}^{2\pi}d\theta e^{-im\theta}e^{i\boldsymbol{A}(\theta)\cdot\boldsymbol{\delta}}$ \cite{Oka_review_2019}, where $\theta = \Omega_p t$ and $\boldsymbol{A}(t)= -\int^t dt' \boldsymbol{E}(t')$.  In our particular case,
\begin{align}
t_{i,i+\boldsymbol{\delta}_{i}}^{\left(m\right)} & =\frac{t_{1}}{2\pi}\int_{0}^{2\pi}d\theta e^{-im\theta}e^{iA_{+}\sin\tilde{\theta}_{+}+iA_{-}\sin\tilde{\theta}_{-}},
\label{eq:t_sums}
\end{align}
where $A_\pm = A_0 (1\pm N^{-1})^{-1}$, with fluence $A_{0}=E_0/(2\Omega)$, and $\tilde{\theta}_{\pm} =\theta\left(N\pm1\right)\mp\phi_{i}$, where $\phi_i$ gives the directional dependence.  The integral can be performed by decomposing both ($\pm$) exponentials into sums over Bessel functions using $\exp (i x \sin \rho)=\sum_{m'} \mathcal{J}_{m'}(x)\mathrm{e}^{i m'\rho}$ \cite{abramowitz1970handbook}, 
\begin{align}
t_{i,i+\boldsymbol{\delta}_{i}}^{\left(m\right)}& =t_1 \!\!\!\!\!\!\!\!\!\sum_{m_{1},m_{2}=-\infty}^{+\infty}\!\!\!\!\!\!\!\!\mathcal{J}_{m_{1}}(A_{+})\mathcal{J}_{m_{2}}(A_{-})e^{-i\left(m_{1}-m_{2}\right)\phi_{l}}\nonumber \\
 & \quad \times\delta_{m-N\left(m_{1}+m_{2}\right)+m_{2}-m_{1}}.\label{eq:t_general_function}
\end{align}
The sums over $m_{1,2}$ can be calculated numerically for any integer $N$. It is convenient to parametrize  $m=N\tilde{m}+k$, with  $\tilde{m} = m_1+m_2$ and $k = m_2-m_1$ integers, which allows the hopping to be written as $t_{i,i+\boldsymbol{\delta}_{l}}^{\left(N\tilde{m}+k\right)} \equiv t_{1} f_{k}^{\tilde{m}}e^{-ik\phi_{l}}$, with $f_{k}^{\tilde{m}}=\mathcal{J}_{\frac{1}{2}(\tilde{m}+k)}(A_{+})\mathcal{J}_{\frac{1}{2}(\tilde{m}-k)}(A_{-})$.
For sufficiently large $N$, the non-negligible amplitudes $f_k^{\tilde{m}}$ are tightly clustered around each $\tilde{m}$, with $k \approx 0$.  An example hopping profile is shown in Fig.~\ref{fig:contributions_t} (a) as function of $m$. 
We can now calculate the nearest-neighbor exchange couplings for each direction, $J_1^{\left(\boldsymbol{\delta}_{l}\right)}$ expanding in the excited energies, $U+(N\tilde{m}+k)\Omega_p$,~\cite{lindgren1974BW,RaoPRB2016,PolkovnikovPRL2016}

\begin{equation}
\!J_{1}^{\left(\boldsymbol{\delta}_{l}\right)}\! = 4\!\sum_{\tilde{m},k}\!\frac{t_{l}^{\left(N\tilde{m}+k\right)}t_{l}^{\left(-N\tilde{m}-k\right)}\!\!\!}{U+(N \tilde{m}+k)\Omega_p}.
\label{eq:J1-quasi-monochromatic}
\end{equation}
The results as a function of fluence, $A_0$ are shown in Fig.~\ref{fig:contributions_t}(b) for fixed frequency and several values of $N$. For small $N$, $J_{1}$ is direction dependent (results are for $\boldsymbol{\delta}_{1}$ on the triangular lattice, with $\phi_1 = \pi/3$). As $N$ increases, $J_{1}$ becomes isotropic, and converges to the average over linearly-polarized monochromatic light. 

We now discuss the limit of large, but finite $N$, where we obtain analytical results. We note that the hoppings, Eq.~(\ref{eq:t_general_function}) are dominated by contributions from $m \approx N \tilde{m}$, allowing the sums to be truncated for $k\ll N$. For large $N$, $A_{+}\approx A_{-}\approx A_{0}$.  As the numerators of Eq.~\eqref{eq:J1-quasi-monochromatic} are dominated by small $k/N$ for each $\tilde{m}$, we approximate
\begin{equation}
\!J_{1}^{\left(\boldsymbol{\delta}_{l}\right)}\! 
\approx 4t_1^2\!\sum_{\tilde{m}}\!\frac{\sum_{k}\!\!\left|f_{k}^{\tilde{m}}\right|^{2}}{U+\tilde{m}\Omega}.
\label{eq:J1-quasi-monochromatic-2}
\end{equation}
Here, we neglect the $k$ dependence of the denominator, but one must be careful, as the $k$ dependence appears to give further resonances at \emph{every} $k$ value. The numerators are strongly suppressed in $k/N$, however, as shown in Fig.~\ref{fig:contributions_t}(a) and so only the resonances near the main $\Omega = -U/\tilde{m}$ resonance are dangerous.  The above result then takes the same form as the magnetic exchange couplings for monochromatic light with \emph{effective} hoppings $t_{1}\sqrt{\sum_k |f_k^{\tilde{m}}|^2}$.  These are independent of $\phi_l$, making $J_{1}^{\left(\boldsymbol{\delta}_{l}\right)}$ isotropic; and $f_k^{\tilde{m}}$ is even with respect to $k$, which guarantees that chiral terms vanish for large $N$\cite{supp}. Chiral fields ($J_\chi$) couple to the scalar chirality $\vec{S}_i\cdot(\vec{S}_j\times \vec{S}_k)$, and are the manifestation of time-reversal symmetry breaking; these may be calculated within fourth order perturbation theory \cite{ClaassenNatComm2017, supp}. The vanishing of chiral terms as $N$ increases is shown in Fig.~\ref{fig:contributions_t}(b).  These analytical results agree well with the exact numerical sums, for $\Omega$ detuned from the resonances and sufficiently large $N \gtrsim 10$. Moreover, they agree with the simple average of the monochromatic Floquet results over all linear polarizations.

In this concrete example, we can address the experimental feasibility of the time scales.  The time for the spins to relax to the new low energy state given by the nonequilibrium exchange couplings is $T_{rel} \sim 1/|J_1|$.  The spins must feel the \emph{unpolarized} exchange couplings, and so $T_{rel} \gg T_p$.  All this, and the measurement must happen within a single laser pulse.  Most generously, we require $T_{pulse} \gg T_{rel} \gg T_p \gg T$, where $T_{pulse}$ is the duration of the pulse. When this hierarchy of time scales is fulfilled, experiments should realize the effective models discussed here, not the time-dependent set of couplings, $J_1^{\left(\boldsymbol{\delta}_{l}\right)}(t)$.  We further discuss the time scales in the supplementary material to argue that these are experimentally plausible \cite{supp}. 


Now we turn to general unpolarized light, where different protocols can lead to different physics. Any polarization profile can be decomposed into Stokes parameters~\cite{bornwolf_book},
\begin{equation}
\vec{S} = I\left(\cos2\chi\cos2\psi ,\cos2\chi\sin2\psi,\sin2\chi\right),\label{eq:V-eq}
\end{equation}
which describe the surface of a sphere of radius $\sqrt{I}$: the Poincar\'e sphere  (Fig.~\ref{fig:Poincar_sphere}), where $I$ is the intensity. For fixed monochromatic light, $\vec{S}$ describes a point on the surface. The poles, $\chi = \pm \pi/4$ correspond to left and right circularly polarized (CP) light, respectively, while linear polarization (LP) lies on the equator ($\chi =0$), with angle $\psi$.  For unpolarized, nearly monochromatic light, the polarization vector slowly traverses a periodic path on the Poincar\'e sphere with characteristic time, $T_p = 2\pi/\Omega_p \gg T = 2\pi/\Omega$, such that the time average of the Stokes parameters is zero, $\langle \vec{S}\rangle =0$ \cite{bornwolf_book,PiqueroOptics2018,ColasLight2015,BeckleyOptExp10,ortega17,shevchenko17,zhu15,shevchenko19,hannonen19}, of which the above case is one example. 
Generically, effective couplings in correlated systems are sensitive to the type of unpolarized light, which can be tuned.  Unpolarized light is differentiated by higher-order correlators of the Stokes parameters, $\langle S_i S_j \rangle$, $\langle S_i S_j S_k \rangle$, etc\cite{klyshko97}, which must also preserve lattice and time-reversal symmetries for the correlated physics to respect those symmetries.  

To preserve lattice and time-reversal symmetries, polarization distributions, $f(\chi,\psi)$ must be invariant under rotations and have zero net chirality. Such distributions generate ``type II'' light \cite{LehnerPRA1996}.  We have already discussed type II Glauber light, which samples all LPs equally, encompassing the equator of the Poincar\'e sphere. Generic type II light may be constructed from superpositions of distributions with circles at $\chi = \pm \chi_0$, $f(\chi,\psi)=\frac{1}{2}\left[\delta(\chi-\chi_{0})+\delta(\chi+\chi_{0})\right]$. Type I light is more restrictive, sampling the Poincar\'{e} sphere uniformly, $f(\chi, \psi)=1$ ~\cite{LehnerPRA1996}. Fixed intensity type I light is known as amplitude-stabilized unpolarized light, while natural light has a varying intensity, $f(I,\chi,\psi)=\frac{2}{I_{0}}\exp\left(-2I/I_{0}\right)$~\cite{goodman2015statistical}; for exchange couplings, these give identical results after averaging over $I$.  It is possible to generate nearly monochromatic type II Glauber ~\cite{ColasLight2015} and type I light~\cite{BeckleyOptExp10,PiqueroOptics2018} either using spatial depolarizers or by superimposing slightly frequency detuned incoherent laser beams with orthogonal polarizations~\cite{supp}.  

Any type of unpolarized light may be explicitly constructed by combining pairs of detuned lasers. The example above used a pair with equal weights of detuned LCP and RCP beams to produce a polarization vector traversing the equator. Any latitude may be traversed using a similar pair with unequal weights, and our analysis can proceed similarly. Different latitudes may then be superimposed by superimposing incoherent pairs of beams\footnote{Coherent superpositions will create more complicated Lissajous figures.}.  Therefore, for any type, we can calculate the couplings for an arbitrary fixed polarization for monochromatic light (see Supplemental Material~\cite{supp}) and simply average over the polarization distribution\footnote{ We fix the intensity, but it may vary, as for natural light~\cite{bornwolf_book}.}, as shown in the previous example. For a given protocol, the magnetic exchange couplings $J_{ij}$ are found by averaging,
\begin{align}
\left\langle J_{ij}\right\rangle = & \frac{\int_{-\pi/4}^{\pi/4}d\chi\,\int_{0}^{\pi}d\psi\,\cos 2\chi\,f(\chi,\psi)J_{ij}(\chi,\psi)}{\int_{-\pi/4}^{\pi/4}d\chi\,\int_{0}^{\pi}d\psi\,\cos2\chi\,\,f(\chi,\psi)}.\label{eq:avg_J-1}
\end{align}

To demonstrate how varying the polarization protocol can drive materials through different regions of phase space, we explicitly consider the triangular lattice.  It provides an apt example, as multiple spin liquids are accessible via different directions in phase space. While the nearest neighbor ($J_1$) model has 120$^\circ$ order, spin liquids may be accessed by adding second neighbor ($J_2$), chiral ($J_\chi$) or ring exchange ($J_\square$) terms.  There is a Dirac spin liquid for $J_2/J_1 \gtrsim 0.1$\cite{hu15,zhu15,li15,iqbal16,gong17,saadatmand17,wietek17,gong_PRB_2019}; a chiral spin liquid for either $J_\chi/J_1 \gtrsim 0.2$ and $J_2 = 0$ or $J_\chi/J_1 \gtrsim 0.025$ for $J_2/J_1 \sim 0.1$\cite{wietek17}; and a spinon Fermi surface state for $J_\square/J_1 \gtrsim 0.2$\cite{Motrunich_PRB_2005}.  The relevant Floquet engineered couplings may be found
by expanding in $U+m\Omega$ either via the Brillouin-Wigner perturbation theory to fourth order \cite{lindgren1974BW,RaoPRB2016}, used in this work, or a Schrieffer-Wolff transformation \cite{PolkovnikovPRL2016}(details in the supplemental material\cite{supp}). Here, we fix the polarization and later average following Eq.~\eqref{eq:avg_J-1} to find the desired unpolarized result.

\begin{figure}

\includegraphics[width=.95\columnwidth]{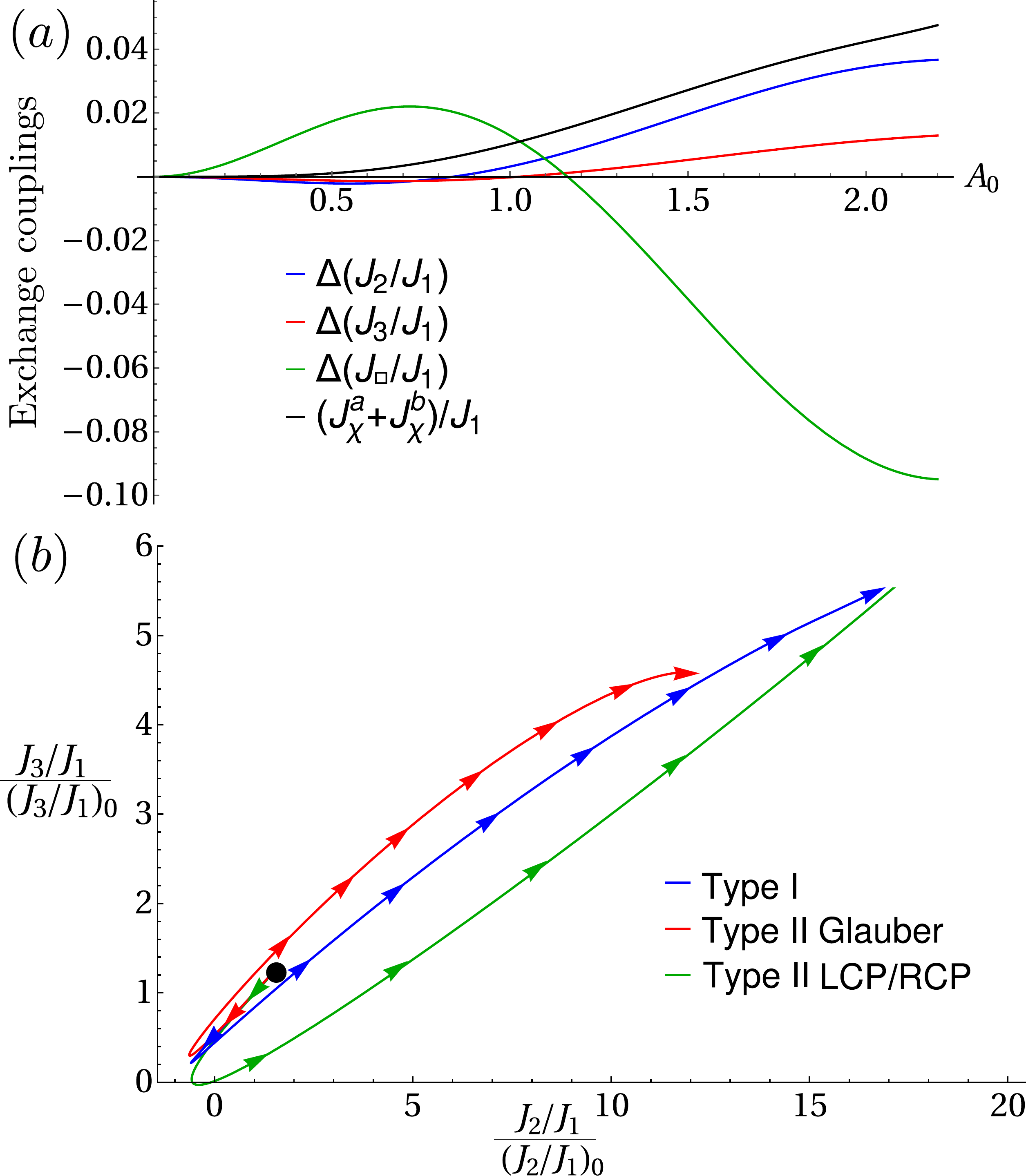}

\caption{Enhancement of magnetic couplings on the triangular lattice with as functions of fluence.  (a) shows the absolute changes for CP light with $\Omega = 2U/3$, where the enhancement is largest.  $J_2/J_1$ and $J_3/J_1$ can be enhanced by $0.03$ and $0.01$, respectively.  These may seem small, but are nearly 2000\% and 500\% of the equilibrium values, as shown in (b), and are a significant fraction of the $J_2/J_1$ required for the Dirac spin liquid.  The effective chiral field reaches $\sim 0.05 J_1$ \cite{supp}, again a significant fraction of the critical field. Ring exchange, $J_\square/J_1$ ranges between $-0.09$ and $0.02$; positive values eventually induce a spin liquid but must be 10x larger. (b) shows how different types of unpolarized light drive different paths through phase space, given in terms of the relative enhancement. A dot indicates the initial equilibrium point ($A_0 = 0$).  Type I light (blue) samples the Poincar\'e sphere evenly; type II Glauber light (red) samples all linearly polarized light equally; and type II LCP/RCP (green) samples only the poles of the Poincar\'e sphere, eliminating chiral fields.  Note that the CP light used in (a) gives identical results to type II LCP/RCP for $J_1$, $J_2$, and $J_3$.
\label{fig:results}}
\end{figure}

To maximally enhance the further neighbor exchange couplings, we must approach the resonances at $\Omega = -U/m$. Yet, if the frequency is too close, doublons and holons are excited and heating is a serious problem.  The  Hubbard bands have a finite bandwidth, $2 \gamma t_1$, where $\gamma$ is lattice dependent  ($\gamma = 2\sqrt{5}$ for the triangular lattice \cite{BalentsPRL2018}), so to avoid heating upon approaching the $U = \Omega$ resonance from below, we keep $\Omega < U-2\gamma t_1$. 
\footnote{Approaching from above ($\Omega > U+2\gamma t_1$) does not lead to substantial enhancements.}
We also must insist, given our fluences, that \emph{two} photons cannot excite electrons between Hubbard bands, $2\Omega > U+2 \gamma t_1$ \footnote{Fortunately, considering $m$ photons does not lead to further restrictions}.  This restriction limits potential materials, as only strongly insulating materials with $t_1 < U/(6\gamma)$ allow strong enhancements without heating.
We fix $t_1 = U/(6\gamma)$ and $\Omega/U = 2/3$ to avoid heating while maximizing the enhancements; see the red vertical line in Fig.~\ref{fig:contributions_t} (a).
Sufficiently far from resonance, there is minimal heating even for large fluences\cite{ClaassenNatComm2017}.
We calculated the enhancements of $J_1$, $J_2$, $J_3$, and $J_\square$ for all kinds of type II and type I light.
$J_2/J_1$ is maximally enhanced by either type I light; type II light with only equal parts LCP and RCP light; or CP light, which also generates $J_\chi$.  We show both the absolute change, Fig. \ref{fig:results}(a) and enhancement over equilibrium values, Fig. \ref{fig:results}(b) as functions of fluence. Due to the Bessel function structure, moderate fluences maximize the enhancement\footnote{we must also take care to remain in the regime where perturbation theory makes sense, as the fourth-order corrections to $J_1$ can potentially drive it negative for some range of fluences.  This is further discussed in the Supplementary Material \cite{supp}}.  The absolute changes can be as large as 25\% and 33\% of the critical $J_\chi/J_1$ and $J_2/J_1$, respectively.  While these will not drive the $t_1$ Hubbard model into a spin liquid, a material with sufficiently large preexisting $J_2$, due either to second neighbor hopping or superexchange, could be tuned to both Dirac and chiral spin liquids via different protocols.  
These absolute changes understate the enhancement, as the equilibrium values are tiny for the $t_1/U$ required to avoid heating, and the enhancement of $J_2/J_1$ can be as large as 2000\%.

Polarization protocols trace out unique paths through the $J_2/J_1 - J_3/J_1$ phase space, as shown in Fig.~\ref{fig:results}(b), where $J_3/J_2$ varies by a factor of two.  Minimizing $J_3$ is essential to access the Dirac spin liquid, as $J_3$ increases the critical $J_2$~\cite{gong_PRB_2019}, and so type I or CP light is more favorable than type II Glauber.  Note that we show two extremes of type II light ($\chi = 0, \pm \pi/4)$, but all type II light lies between these.  

We have shown that unpolarized light provides an untapped tuning parameter for Floquet engineering, and possibly nonequilibrium physics in general, particularly for correlated materials sensitive to higher-order correlations in the polarization. We showed that calculations can be done using Floquet techniques with fixed polarization and then averaged, as long as the polarization vector varies sufficiently slowly ($T_p \gtrsim 10 T$). We illustrated this effect on magnetic exchange couplings for the triangular lattice and showed how different types of unpolarized light drive the model through varied directions in phase space.  In particular, the same $J_1-J_2$ triangular material could be nudged into either Dirac or chiral spin liquids by different polarization protocols.  Similar effects should be found throughout correlated materials. Future research might examine incommensurate frequencies, where the pseudorandom nature of the polarization variation may have interesting effects.

We acknowledge useful discussions with Thomas Iadecola, Eduardo Miranda, Peter Orth, Paraj Titum, Thais Trevisan, Chirag Vaswani, and Jigang Wang. V.L.Q and R.F. and were supported by NSF through grant DMR-1555163. RF thanks the Aspen Center for Physics, supported by the NSF Grant PHY-1607611, for hospitality.

\begin{widetext}
\section{Supplemental material}

\subsection{Monochromatic light}

We revisit the Floquet formalism for monochromatic light. Mochromatic light is fully polarized, with the electric field delineating an ellipse perpendicular to the propagation vector, $\vec{E}(t) = \mathrm{Re}\left[\vec{E}_0 \mathrm{e}^{-i \Omega t}\right]$, $\vec{E}_0$ is independent of time.  We consider propagation along $\hat z$, normal to the sample.  A generic polarization can be written as $\vec{E}_0 = E_+ \hat \epsilon_+ + E_- \hat \epsilon_-$, where $\hat \epsilon_\pm = \frac{1}{\sqrt{2}}(\hat x \pm i \hat y)$ are left and right circular polarization (LCP/RCP).  $E_\pm$ is characterized by its amplitude $\sqrt{I}$ and angles $\chi \in (-\pi/4,\pi/4)$ and $\psi \in (0,\pi)$, $E_{\pm} =\sqrt{I}\sin\left(-\chi\mp\pi/4\right)e^{\mp i\left(\psi-\pi/2\right)}$. 
The double occupancy penalty becomes $U+m\Omega$, with resonances at $m = - \Omega/U$.  The hopping between sectors is given by the Fourier transform (with $\theta = \Omega t$), \cite{AokiPRB2016}
\begin{align}
t_{i,i+\boldsymbol{\delta}_{i}}^{\left(m\right)}& =\frac{t_{1}}{2\pi}\int_{0}^{2\pi}d\theta e^{-im\theta}e^{-i\boldsymbol{\delta}_{i}\cdot\boldsymbol{A}\left(\theta\right)}=t_{1}e^{im\left(\beta_{l}+\pi\right)}\mathcal{J}_{m}\left(A_{l}\right),\label{eq:hops-final}
\end{align}
where the Bessel function $\mathcal{J}_{m}$ encodes the real space orientation $\boldsymbol{\delta}_{l}$ via the amplitude, $A_l$ and angle, $\beta_l$,
\begin{align}
A_{l} & =A_{0}\sqrt{1+\cos2\chi\cos\left[2\left(\psi-\phi_{l}\right)\right]}\cr
\cos\beta_{l} & =\frac{\sqrt{2}\sin\chi\sin\left(\psi-\phi_{l}\right)}{\sqrt{1+\cos2\chi\cos\left[2\left(\psi-\phi_{l}\right)\right]}}.\label{eq:cos-bl}
\end{align}
Here, we introduce the dimensionless fluence $A_{0}=\frac{1}{\Omega}\sqrt{I/2}$.
Notice that $A_{l}$ is symmetric with respect to $\chi=0$, while $\beta_l \rightarrow \pi - \beta_l$ as $\chi \rightarrow - \chi$, which explains the lack of time-reversal symmetry breaking in distributions that sample $\pm \chi$ equally. Now one can calculate the exchange couplings, with modified hoppings and $U+m \Omega$ denominators. The nearest-neighbor coupling was extensively explored before~\cite{PolkovnikovPRL2016,ClaassenNatComm2017,BalentsPRL2018,BalentsPRB2018} , 
\begin{equation}
J_{1}^{\left(\boldsymbol{\delta}_{l}\right)} = 4 \sum_m \frac{t_{i,i+\boldsymbol{\delta}_{l}}^{(m)} t_{i+\boldsymbol{\delta}_{l},i}^{(-m)}}{U+m \Omega} = 4t_{1}^{2}\sum_{m}\frac{\left|\mathcal{J}_{m}\left(A_{l}\right)\right|^{2}}{U+m\Omega}.\label{eq:J1-order-2}
\end{equation}
 The Bessel functions cause $J_1^{\left(\boldsymbol{\delta}_{l}\right)}$ to rise to a maximum as a function of fluence, $A_0$ and then oscillate with a decaying envelope. The anisotropy of the lattice is generically unavoidable given the dependence of $A_l$ on the hopping direction. Higher-order contributions are more complicated and lattice-dependent, as superexchange paths proliferate; third-order terms vanish, while fourth-order terms on the triangular lattice are derived in the next Sections. Imaginary hopping terms, if present, generate chiral fields, $J^\triangle_\chi \sum_{ijk \in \triangle}\vec{S}_i\cdot \vec{S}_j \times \vec{S}_k$.  
Otherwise, the corrections modify existing couplings.

\subsection{Time scales and experimental details}

In this section, we discuss the different time scales, frequencies and fluences involved, and discuss experimental feasibility.  Here, our degrees of freedom are spins, with interaction scale $J_1 = 4 t_1^2/U$, that are experiencing a pulse of light (duration, $T_{pulse}$) of frequency $\Omega = 2\pi/T$.  We assume that the polarization vector oscillates with period $T_p \gtrsim 10 T$, such that polarization averaging is expected to be reasonable.  

In order to maximally enhance the exchange couplings, $\Omega = 2 U/3$, and $t_1 = U/(6\gamma)$, where $\gamma = 2 \sqrt{5}$ for the triangular lattice.  These time scales can be well separated, with perhaps the most stringent requirement being for the pulse length required to allow the spins to relax,
\begin{equation}
    T \sim \frac{1}{U} \ll T_p \sim \frac{10}{U} \ll T_{rel} \sim \frac{100}{U} \ll T_{pulse}.
\end{equation}

These laser frequencies will need to be tuned to the Mott gap, and so are expected to be on the order of electron volts, in the visible range.  $T_p$ will therefore be on the order of $10$fs, while $T_{rel} \sim 100$fs, requiring a moderately long pulse.  Note that here we consider only how photons affect the electronic degrees of freedom in this single-band Hubbard space directly. In general, real materials will have additional spin relaxation channels, as photons interact with collective modes, like phonons or spin waves, or excite electrons into other bands \cite{Jigang_Nature_2013}; these details will be materials specific.


The dimensionless vector potential amplitude can be estimated by restoring
the units,
\begin{equation}
A_{0}=\frac{a_{0}eE}{\Omega\hbar},
\end{equation}
where $a_{0}$ is the lattice spacing, of the order of Angstroms.
This amplitude is connected to
intensity, with full units, according to
\begin{align}
I & =c\epsilon_{0}\left(\frac{\Omega\hbar}{ea_{0}}\right)^{2}\left|A_{0}\right|^{2}=2.6\times10^{17}\left(\frac{\Omega\hbar\left[eV\right]}{a_{0}\left[\AA\right]}\right)^{2}\left|A_{0}\right|^{2}W/m^{2}
\end{align}
with $\epsilon_{0}$ the vacuum permittivity. The electric field strength, $eE$
varies in different experiments, typically ranging from $\left(0.01-1\right)eV/\AA\,$\cite{Huber2008OptLett,Wang2013Science},
giving to intensities of $I\approx10^{15}-10^{17}W/m^{2}$. In these experiments, $A_0$ ranges between $0.01$ and $1$; the slightly larger values of $\sim 2$ that we require are not unreasonable.  However, as lasers provide constant power that can be chopped into pulses, either shorter pulses with larger fluences, or longer pulses with lower fluences, at the moment the two requirements of relatively high fluence and relatively long pulses are at odds, given current technology.

In addition to driving the system into a nonequilibrium state, the state itself must be measured via some optical measurements. Ordered phases should be more or less straightforward, as a phase transition should give a clear signal in optical quantities. However, we propose to drive materials into spin liquid regions that do not exist in equilibrium materials.  Here, the absence of a phase transition would just be the minimal requirement for realizing a spin liquid.  Electromagnetic gauge fields do interact with the neutral spinons, albeit often with significantly lower amplitudes than electrons.  Gapless spin liquids are predicted to have power-law behavior of the optical conductivity \cite{potter13}, with some evidence in herbertsmithite and others~\cite{pilon13,pustogow18}, and spin liquids may have signatures in the magneto-optical Faraday or Kerr effects \cite{colbert14}.

One alternative to averaging the polarization over time is to average the polarization \emph{spatially}.  A uniformly polarized beam may be passed through an optical element called a depolarizer that causes the polarization to vary spatially such that a spatial average has zero net polarization, $\langle \vec{S}\rangle = 0$ \cite{burns1983Lightwave,mcguire90,hodgson2005laser}. Current depolarizers  have length scales significantly larger than the lattice spacing, and care must be taken that the material experiences a polarization average, not just spatially disordered couplings. Practically speaking, the Brownian motion of the laser beam might be exploited to effectively randomize the exchange couplings on any given link and thus average the polarizations.

\subsection{Definition of the magnetic exchange couplings}

In this section, we define the exchange couplings of the effective
spin Hamiltonian for the triangular lattice.
The nearest-neighbor vectors are given by 
\begin{equation}
\boldsymbol{\delta}_{1}=\left(1/2,\sqrt{3}/2\right),\,\,\boldsymbol{\delta}_{2}=\left(1,0\right),\,\,\boldsymbol{\delta}_{3}=\left(1/2,-\sqrt{3}/2\right).
\end{equation}
The distinct exchange terms are shown in Fig.~\ref{fig:all-paths-triangular} yielding the Hamiltonian

\begin{align}
H_{{\rm spin}}^{\left(m\Omega\lessapprox U\right)} & =\sum_{\left\langle i,j\right\rangle }J_{1}^{\left(i,j\right)}\boldsymbol{S}_{i}\cdot\boldsymbol{S}_{j}+\sum_{\left\langle \left\langle i,k\right\rangle \right\rangle }J_{2}^{\left(i,k\right)}\boldsymbol{S}_{i}\cdot\boldsymbol{S}_{j}+\sum_{\triangle}J_{\chi}^{\left(i,j,k\right)}\chi_{\triangle}^{\left(i,j,k\right)}+\sum_{\left\langle \left\langle \left\langle i,m\right\rangle \right\rangle \right\rangle }J_{3}^{\left(i,m\right)}\boldsymbol{S}_{i}\cdot\boldsymbol{S}_{m}+\nonumber \\
 & +\sum_{\square}\left[J_{\square}^{\left(i,j,k,l\right)}P_{\square}^{\left(i,j,k,l\right)}+J_{\square}^{\left(i,l,j,k\right)}P_{\square}^{\left(i,l,j,k\right)}\right. \left.-J_{\square}^{\left(i,k,j,l\right)}P_{\square}^{\left(i,k,j,l\right)}\right].\label{eq:Hamil_square}
\end{align}
 The couplings $J_{1}$, $J_{2}$, and $J_{3}$ are the nearest, next-nearest, and third-neighbor couplings. $J_\square$'s are the ring exchange terms that, in our notation, multiply the 4-body operators

\begin{equation}
P_{\square}^{\left(i,j,k,l\right)}=\left(\boldsymbol{S}_{i}\cdot\boldsymbol{S}_{j}\right)\left(\boldsymbol{S}_{k}\cdot\boldsymbol{S}_{l}\right),\label{eq:P_ring}
\end{equation}
the product of all the spin operators around a given plaquette. For
any choice of polarization average that keeps the lattice symmetries,

\begin{equation}
J_{\square}^{\left(i,j,k,l\right)}=J_{\square}^{\left(i,l,j,k\right)}=J_{\square}^{\left(i,k,j,l\right)}.
\end{equation}
For the time-independent case, to second order, there is only the nearest-neighbor term \cite{mahan2010book}, $J_1 = 4t_1^2/U$, but fourth order terms give corrections to $J_1 = 4t_1^2/U-28 t_1^4/U^3$ \cite{MacDonaldPRB1988}, as well as $J_2 = J_3 = 4t_1^4/U^3$ and $J_\square = 80 t_1^4/U$~\cite{MacDonaldPRB1988,Tremblay2004PRB}.

The chiral couplings come in two flavors, shown in Fig.~\ref{fig:all-paths-triangular} (d) and (e). In (e), the electron hops around a closed lattice triangle, while in (d), the three sites form an open path. We call $J_{\chi}^{a}$ the processes coming from (d) and $J_{\chi}^{b}$ the ones coming from (e). It becomes natural to find the net chirality of a triangle, by distributing the different fluxes coming from the two terms. Considering four sites forming a parallelogram, like the one shown in (b),  the net flux consists of adding two fluxes of (d) and two fluxed of (e). This parallelogram is made of two triangles, implying that the coupling that controls the effective chirality is $J_{\chi}^{a}+J_{\chi}^{b}$. This is used as the reduced variable in the main text.

\begin{figure}[htbp]
\includegraphics[width=1.0\columnwidth]{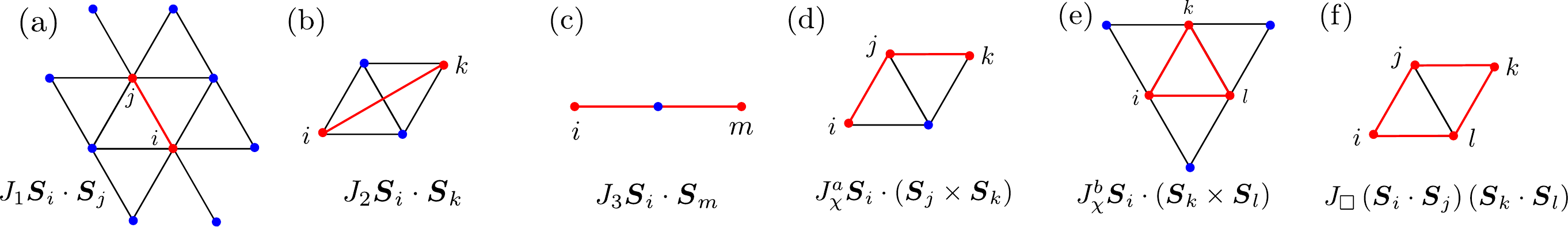}\caption{Representation of all the sites involved that lead to exchange couplings in fourth order in $t$ on the triangular lattice, with the bond sites represented in red. (a) $J_{1}$ (b) $J_{2}$ (c) $J_{3}$ (d) $J_{\chi}^{a}$
(e) $J_{\chi}^{b}$ (f) $J_{\square}$. \label{fig:all-paths-triangular}}
\end{figure}

\subsection{Magnetic exchange couplings on the triangular lattice}

We now present the main features of the perturbative expansion leading to the effective magnetic exchange couplings shown in the main text; an expanded calculation will be shown elsewhere \cite{QuitoFlintPRB2020}.  This calculation can be done two ways, following the Brillouin-Wigner~\cite{lindgren1974BW} or Schrieffer-Wolff \cite{SchriefferWolfforiginal,PolkovnikovPRL2016}. We take the Brillouin-Wigner approach here.

The Hilbert space of the problem is enlarged when the Floquet modes
are introduced. The identity operator in the full Hilbert space formed
by joining the Floquet and Fock spaces reads 
\begin{align}
\mathds{1} & =\mathds{1}_{\text{Fock}}\otimes \mathds{1}_{\text{\text{Floquet}}}\equiv\mathcal{P}+\mathcal{Q},\label{eq:identity_full}
\end{align}
with $\mathcal{P}$ and $\mathcal{Q}$ the projectors onto the ground
state and excited states manifolds of the \emph{full} Floquet-Fock Hilbert
space. The total ground state projector $\mathcal{P}$ is the tensor product of the Fock and Floquet ground state manifolds,
$\mathcal{P}=P\otimes P_{F,0}$, while the projector onto excited
states is

\begin{equation}
\mathcal{Q}=\sum_{m=-\infty}^{+\infty}P_{F,m}Q+\sum_{m\ne0}P_{F,m}P,
\end{equation}
with $Q$ the excited states of the fermions, only. 

When compared to the time-independent case, the novel effects in the structure of the perturbation theory comes from the second term of $\mathcal{Q}$, which projects
onto the fermionic ground state manifold as long as $m\ne0$ in Floquet
space. 

The resolvent operator $\mathcal{R}$, which encodes the sum over the excited states and takes into account the energy denominators is $\mathcal{R}=\mathcal{R}_{1}+\mathcal{R}_{2}$, where

\begin{align}
\mathcal{R}_{1} & =\frac{\sum_{m}P_{F,m}Q}{E_{0}-\mathcal{H}_{0}},\label{eq:R_1}\\
\mathcal{R}_{2} & =\frac{\sum_{m\ne0}P_{F,m}P}{E_{0}-\mathcal{H}_{0}},\label{eq:R_2}
\end{align}
with $E_{0}$ is the ground state energy of $\mathcal{H}_{0}$.

The information coming from the hopping Hamiltonian is used to construct
the wave operator $\mathcal{W}$, which is implicitly defined
by~\cite{lindgren1974BW}

\begin{equation}
\mathcal{W}=\mathcal{P}+\mathcal{R}\left(\mathcal{V}\mathcal{W}-\mathcal{W}\mathcal{V}\mathcal{W}\right).\label{eq:W-eq}
\end{equation}
The effective spin Hamiltonian is obtained from $\mathcal{W}$

\begin{align}
H_{{\rm spin}}^{\left(m\Omega\lessapprox U\right)} & =\mathcal{P}\mathcal{H}_{0}\mathcal{P}+\mathcal{P}\mathcal{V}\mathcal{W}=\mathcal{P}\mathcal{V}\mathcal{W},\label{eq:effect-Hamilt}
\end{align}
where the second equality follows given that the projection of $\mathcal{H}_{0}$
onto the ground state manifold is zero. The equation
for the wave operator can be solved recursively to a certain order
of the perturbation potential $\mathcal{V}$.
The zeroth order term from Eq.~(\ref{eq:W-eq}) to $\mathcal{W}$
is $\mathcal{W}^{\left(0\right)}=\mathcal{P}$~\cite{lindgren1974BW}. This term gives a vanishing contribution to the effective Hamiltonian
Eq.~(\ref{eq:effect-Hamilt}) given that $\mathcal{P}$ projects
onto the Fock ground state with one electron per site while $\mathcal{V}$
moves electrons creating empty and doubly occupied states. Similar
reasoning leads to the conclusion that all terms with an even number
of $\mathcal{V}$ insertions in $\mathcal{W}$ will also vanish. The
leading contributions to $\mathcal{W}$ are found from first and third
order in $\mathcal{V}$~\cite{lindgren1974BW},

\begin{align}
\mathcal{W}^{\left(1\right)} & =\mathcal{R}\mathcal{V}\mathcal{P},\\
\mathcal{W}^{\left(3\right)} & =\mathcal{R}\mathcal{V}\mathcal{R}\mathcal{V}\mathcal{R}\mathcal{V}\mathcal{P}-\mathcal{R}^{2}\mathcal{V}\mathcal{P}\mathcal{V}\mathcal{R}\mathcal{V}\mathcal{P}.
\end{align}
From Eq.~(\ref{eq:effect-Hamilt}), $\mathcal{W}^{\left(1\right)}$
and $\mathcal{W}^{\left(3\right)}$ lead to the effective spin Hamiltonians in orders two and four,

\begin{align}
\mathcal{\mathcal{H}}^{\left(2\right)} & =\mathcal{P}\mathcal{V}\mathcal{R}\mathcal{V}\mathcal{P},\label{eq:H-spins-2nd-order}\\
\mathcal{\mathcal{H}}^{\left(4\right)} & =\mathcal{P}\mathcal{V}\mathcal{R}\mathcal{V}\mathcal{R}\mathcal{V}\mathcal{R}\mathcal{V}\mathcal{P}-\left(\mathcal{P}\mathcal{V}\mathcal{R}^{2}\mathcal{V}\mathcal{P}\right)\mathcal{\mathcal{H}}^{\left(2\right)}.\label{eq:H-spins-4th-order}
\end{align}

\subsubsection{Second-order perturbation theory}

The second-order correction $\mathcal{\mathcal{H}}^{\left(2\right)}$
can be calculated by decomposing $\mathcal{R}$ as the sum of $\mathcal{R}_{1}$
and $\mathcal{R}_{2}$ and noticing, from Eq.~(\ref{eq:R_2}), that
$\mathcal{R}_{2}\mathcal{V}\mathcal{P}=0$ since, explained earlier,
$\mathcal{P}\mathcal{V}\mathcal{P}=0$ . In second-order perturbation theory, therefore,
$\mathcal{R}_{2}$ does not enter the calculation and the structure
is identical to the time-independent model, except for the energy
denominators and renormalized hoppings. By plugging the resolvent
$\mathcal{R}_{1}$ explicitly, Eq.~(\ref{eq:H-spins-2nd-order}), and defining $
\mathcal{V}_{m_{1}-m_{2}}=\left\langle m_{1}\left|\mathcal{V}\right|m_{2}\right\rangle $
we arrive at

\begin{align}
\mathcal{\mathcal{H}}^{\left(2\right)} & =-\sum_{m}\left(P\mathcal{V}_{m}Q\right)\frac{1}{\left(U+m\Omega\right)}\left(Q\mathcal{V}_{-m}P\right).\label{eq:second-order-H}
\end{align}

By inserting $\mathcal{V}_{m}$, we arrive at Eq.~(\ref{eq:J1-order-2}).

\subsubsection{Third-order perturbation theory}
Although the Floquet fields break time-reversal symmetry dynamically, the contributions in third-order perturbation theory sum out to zero, including the chiral terms. This is true for any choice of polarization and was previously addressed for circularly polarized light~\cite{ClaassenNatComm2017}.  

\subsubsection{Fourth-order perturbation theory}

Since the third-order corrections vanish, we now proceed to fourth-order. By plugging the resolvent $\mathcal{R}$
into Eq.~(\ref{eq:H-spins-4th-order}), we find that the first term
leads to two possible intermediate steps, with either $\mathcal{R}_{1}$
or $\mathcal{R}_{2}$ in the middle. By separating all the
contributions, we arrive at

\begin{align}
\mathcal{\mathcal{H}}^{\left(4\right)} & =\mathcal{P}\mathcal{V}\mathcal{R}_{1}\mathcal{V}\mathcal{R}_{1}\mathcal{V}\mathcal{R}_{1}\mathcal{V}\mathcal{P}+\mathcal{P}\mathcal{V}\mathcal{R}_{1}\mathcal{V}\mathcal{R}_{2}\mathcal{V}\mathcal{R}_{1}\mathcal{V}\mathcal{P}-\left(\mathcal{P}\mathcal{V}\mathcal{R}_{1}^{2}\mathcal{V}\mathcal{P}\right)\mathcal{\mathcal{H}}^{\left(2\right)}.\label{eq:order-4-generic}
\end{align}
After using equations (\ref{eq:R_1}) and (\ref{eq:R_2}) for
the resolvent, the Hilbert space of the problem is again the Fock
space of the fermions, as only the projectors $P$ and $Q$ are left
in the calculation. By plugging them explicitly, we arrive at

\begin{align}
\mathcal{\mathcal{H}}_{a}^{\left(4\right)}= & -\!\!\!\!\!\sum_{m_{1},m_{2},m_{3}}\frac{P\mathcal{V}_{-m_{3}}Q_{U}\mathcal{V}_{m_{3}-m_{2}}Q_{U}\mathcal{V}_{m_{2}-m_{1}}Q_{U}\mathcal{V}_{m_{1}}P}{\left(U+m_{3}\Omega\right)\left(U+m_{2}\Omega\right)\left(U+m_{1}\Omega\right)}-\!\!\!\!\!\!\sum_{m_{1},m_{2},m_{3}}\frac{P\mathcal{V}_{-m_{3}}Q_{U}\mathcal{V}_{m_{3}-m_{2}}Q_{2U}\mathcal{V}_{m_{2}-m_{1}}Q_{U}\mathcal{V}_{m_{1}}P}{\left(U+m_{3}\Omega\right)\left(2U+m_{2}\Omega\right)\left(U+m_{1}\Omega\right)}\label{eq:H4a}\\
\mathcal{\mathcal{H}}_{b}^{\left(4\right)}=- & \!\!\!\!\!\sum_{m_{1},m_{2}\ne0,m_{3}}\!\frac{P\mathcal{V}_{-m_{3}}Q_{U}\mathcal{V}_{m_{3}-m_{2}}P\mathcal{V}_{m_{2}-m_{1}}Q_{U}\mathcal{V}_{m_{1}}P}{\left(U+m_{3}\Omega\right)\left(m_{2}\Omega\right)\left(U+m_{1}\Omega\right)},\label{eq:H4b}\\
\mathcal{\mathcal{H}}_{c}^{\left(4\right)}= & \sum_{m_{1},m_{2}}\frac{P\mathcal{V}_{-m_{2}}Q_{U}\mathcal{V}_{m_{2}}P\mathcal{V}_{-m_{1}}Q_{U}\mathcal{V}_{m_{1}}P}{\left(U+m_{2}\Omega\right)^{2}\left(U+m_{1}\Omega\right)}.\label{eq:H4c}
\end{align}

In the proceeding equations, we decomposed $Q$ as 

\begin{equation}
Q=Q_{U}+Q_{2U}+Q_{3U}+\ldots,\label{eq:Q_decomp}
\end{equation}
with $Q_{kU}$ projecting onto the fermionic manifold of energy $kU$.

For the explicit calculation of all the couplings that appear from Eqs.~(\ref{eq:H4a})-(\ref{eq:H4c}) for the triangular lattice, it is a matter of summing over over all possible paths. For notation, we refer again to Fig.~\ref{fig:all-paths-triangular}. The effective magnetic
exchange couplings are expressed in terms of the functions $A_{l}$, defined in Eq.~(\ref{eq:cos-bl}), and we define $\tilde{t} = t_1/U$ and $\tilde{\Omega} = \Omega/U$, for simplicity.

\begin{align}
\mathcal{A}_{ijkl}\left(\boldsymbol{m}\right) & =\left(-1\right)^{m_{2}}\tilde{t}^{3}\frac{\mathcal{J}_{-m_{3}}\left(A_{l_{i}}\right)\mathcal{J}_{m_{3}-m_{2}}\left(A_{l_{j}}\right)\mathcal{J}_{m_{2}-m_{1}}\left(A_{l_{k}}\right)\mathcal{J}_{m_{1}}\left(A_{l_{l}}\right)}{\left(1+m_{1}\tilde{\Omega}\right)\left(1+m_{2}\tilde{\Omega}\right)\left(1+m_{3}\tilde{\Omega}\right)},\label{eq:A_m}\\
\mathcal{L}_{ijkl}\left(\boldsymbol{m}\right) & =(-1)^{m_{1}+m_{3}}\tilde{t}^{3}\cos^{2}\left(m_{2}\frac{\pi}{2}\right)\frac{\mathcal{J}_{-m_{3}}\left(A_{l_{i}}\right)\mathcal{J}_{m_{3}-m_{2}}\left(A_{l_{j}}\right)\mathcal{J}_{m_{2}-m_{1}}\left(A_{l_{k}}\right)\mathcal{J}_{m_{1}}\left(A_{l_{l}}\right)}{\left(1+m_{1}\tilde{\Omega}\right)\left(2+m_{2}\tilde{\Omega}\right)\left(1+m_{3}\tilde{\Omega}\right)},\label{eq:L_m}\\
\mathcal{B}_{ij}\left(\boldsymbol{m}\right) & =\left(-1\right)^{m_{1}+m_{3}}\tilde{t}^{3}\cos^{2}\left(m_{2}\frac{\pi}{2}\right)\frac{\mathcal{J}_{-m_{3}}\left(A_{l_{i}}\right)\mathcal{J}_{m_{3}-m_{2}}\left(A_{l_{i}}\right)\mathcal{J}_{m_{2}-m_{1}}\left(A_{l_{j}}\right)\mathcal{J}_{m_{1}}\left(A_{l_{j}}\right)}{\left(1+m_{1}\tilde{\Omega}\right)\left(m_{2}\tilde{\Omega}\right)\left(1+m_{3}\tilde{\Omega}\right)},\,\,m_{2}\ne0,\label{eq:B_m}\\
\mathcal{G}_{ij}\left(\boldsymbol{m}\right) & =\tilde{t}^{3}\delta_{m_{2},0}\left[\mathcal{J}_{m_{1}}^{2}\left(A_{l_{i}}\right)\mathcal{J}_{m_{3}}^{2}\left(A_{l_{j}}\right)+\mathcal{J}_{m_{1}}^{2}\left(A_{l_{j}}\right)\mathcal{J}_{m_{3}}^{2}\left(A_{l_{i}}\right)\right]\frac{1}{\left(1+m_{1}\tilde{\Omega}\right)^{2}\left(1+m_{3}\tilde{\Omega}\right)}.\label{eq:G_m}
\end{align}
where we define $\boldsymbol{m}\equiv\left(m_{1},m_{2},m_{3}\right)$. 

The next-nearest neighbor coupling $J_{2}$ {[}Fig.~\ref{fig:all-paths-triangular}(b){]}
reads

\begin{align}
J_{2}^{\left(i,k\right)} & =\sum_{\boldsymbol{m}}-8\left\{ \mathcal{A}_{1,2,2,1}\left(\boldsymbol{m}\right)\cos^{2}\left[\left(m_{1}+m_{3}\right)\frac{\pi}{2}\right]\cos\left[\left(\beta_{1}-\beta_{0}\right)\left(m_{1}-m_{3}\right)\right]+\mathcal{A}_{1,2,1,2}\left(\boldsymbol{m}\right)\cos^{2}\left[\left(m_{1}+m_{2}+m_{3}\right)\frac{\pi}{2}\right]\right.\times\nonumber \\
\times & \left.\cos\left[\left(m_{1}-m_{2}+m_{3}\right)\left(\beta_{1}-\beta_{0}\right)\right]\right\} +8\mathcal{L}_{2,2,1,1}\left(\boldsymbol{m}\right)\cos\left[m_{2}\left(\beta_{1}-\beta_{0}\right)\right]-16\mathcal{B}_{2,1}\left(\boldsymbol{m}\right)\cos\left[\left(\beta_{1}-\beta_{0}\right)m_{2}\right]+8\mathcal{G}_{2,1}\left(\boldsymbol{m}\right),\label{eq:J2_generic}
\end{align}
while the plaquette terms {[}Fig.~\ref{fig:all-paths-triangular}(f){]}
reads

\begin{align}
J_{\square}^{\left(i,j,k,l\right)}= & \sum_{\boldsymbol{m}}32\left\{ \mathcal{A}_{1,2,2,1}\left(\boldsymbol{m}\right)\cos^{2}\left[\left(m_{1}+m_{3}\right)\frac{\pi}{2}\right]\cos\left[\left(\beta_{1}-\beta_{0}\right)\left(m_{1}-m_{3}\right)\right]+\mathcal{A}_{1,2,1,2}\left(\boldsymbol{m}\right)\cos^{2}\left[\left(m_{1}+m_{2}+m_{3}\right)\frac{\pi}{2}\right]\right.\times\nonumber \\
\times & \left.\cos\left[\left(m_{1}-m_{2}+m_{3}\right)\left(\beta_{1}-\beta_{0}\right)\right]\right\} +32\cos\left[m_{2}\left(\beta_{1}-\beta_{0}\right)\right]\mathcal{L}_{2,2,1,1}\left(\boldsymbol{m}\right).\label{eq:Jsquare_generic}
\end{align}
The $J_{3}$ coupling {[}Fig.~\ref{fig:all-paths-triangular}(c){]}
is

\begin{align}
J_{3}^{\left(i,l,m\right)}= & \sum_{\boldsymbol{m}}-4\mathcal{A}_{2,2,2,2}\left(\boldsymbol{m}\right)+8\mathcal{B}_{2,2}\left(\boldsymbol{m}\right)+4\mathcal{G}_{2,2}\left(\boldsymbol{m}\right).\label{eq:J3_generic}
\end{align}

The chiral term reads {[}Fig.~\ref{fig:all-paths-triangular}(d){]}

\begin{equation}
J_{\chi}^{a\left(i,j,k\right)}=\sum_{\boldsymbol{m}}16\left[\mathcal{L}_{2,2,1,1}\left(\boldsymbol{m}\right)-\mathcal{B}_{2,1}\left(\boldsymbol{m}\right)\right]\left(\sin\left[m_{2}\left(\beta_{1}-\beta_{0}\right)\right]-\sin\left[m_{2}\left(\beta_{1}-\beta_{3}\right)\right]+\sin\left[m_{2}\left(\beta_{2}-\beta_{3}\right)\right]\right).\label{eq:J_chi_generic}
\end{equation}
It might be surprising, at first sight, that the expression of (\ref{eq:J_chi_generic})
has only terms proportional to $\mathcal{L}$ and $\mathcal{B}$,
with the terms proportional to $\mathcal{A}$ vanishing exactly. An
interesting sanity check that this is the case consists of expanding
(\ref{eq:J_chi_generic}) in powers of $1/\Omega$, assuming $\Omega\gg U$.
The leading contribution comes from $1/\Omega^{3}$ and not $1/\Omega$,
as would be naively expected. This is in agreement with the limit of high $\Omega$, where $1/\Omega$ corrections to the
hoppings on the triangular lattice vanishes. $J_{\chi}^{b}$ {[}Fig.~\ref{fig:all-paths-triangular}(e){]}
gives

\begin{equation}
J_{\chi}^{b}=-3J_{\chi}^{a}. \label{eq:J_chi_b}
\end{equation}

These results are generic for light of arbitrary \emph{fixed} polarization. In the main text, we address the vanishing of the chiral terms for the light profile shown in Fig.~2, which presents a slowly varying periodic polarization. This requires generalizing the above expressions for hoppings that do not follow Eq.~\eqref{eq:hops-final}, but instead, Eq.~5. For $J_{\chi}^{b}$ it reads

\begin{equation}
J_{\chi}^{b}=\left[\sum_{m_{1},m_{2},m_{3}}\frac{1}{\left(1+m_{1}\tilde{\Omega}\right)\left(2+m_{2}\tilde{\Omega}\right)\left(1+m_{3}\tilde{\Omega}\right)}-\sum_{m_{1},m_{2}\ne0,m_{3}}\frac{1}{\left(1+m_{1}\tilde{\Omega}\right)\left(m_{2}\tilde{\Omega}\right)\left(1+m_{3}\tilde{\Omega}\right)}\right]g\left(\boldsymbol{m}\right)
\end{equation}
with
\begin{align}
g= & \left[t_{0}^{m_{1}}\left(t_{0}^{m_{1}-m_{2}}\right){}^{*}+\left(t_{0}^{-m_{1}}\right){}^{*}t_{0}^{m_{2}-m_{1}}\right]\left[\left(t_{\frac{\pi}{3}}^{m_{2}-m_{3}}\right){}^{*}t_{\frac{\pi}{3}}^{-m_{3}}-\left(t_{-\frac{\pi}{3}}^{m_{2}-m_{3}}\right){}^{*}t_{-\frac{\pi}{3}}^{-m_{3}}+t_{\frac{\pi}{3}}^{m_{3}-m_{2}}\left(t_{\frac{\pi}{3}}^{m_{3}}\right){}^{*}-t_{-\frac{\pi}{3}}^{m_{3}-m_{2}}\left(t_{-\frac{\pi}{3}}^{m_{3}}\right){}^{*}\right]\nonumber \\
+ & \left[t_{-\frac{\pi}{3}}^{m_{1}}\left(t_{-\frac{\pi}{3}}^{m_{1}-m_{2}}\right){}^{*}+\left(t_{-\frac{\pi}{3}}^{-m_{1}}\right){}^{*}t_{-\frac{\pi}{3}}^{m_{2}-m_{1}}-\left(t_{\frac{\pi}{3}}^{-m_{1}}\right){}^{*}t_{\frac{\pi}{3}}^{m_{2}-m_{1}}-t_{\frac{\pi}{3}}^{m_{1}}\left(t_{\frac{\pi}{3}}^{m_{1}-m_{2}}\right){}^{*}\right]\left[\left(t_{0}^{m_{2}-m_{3}}\right){}^{*}t_{0}^{-m_{3}}+t_{0}^{m_{3}-m_{2}}\left(t_{0}^{m_{3}}\right){}^{*}\right]\nonumber \\
+ & \left[-t_{-\frac{\pi}{3}}^{m_{1}}\left(t_{-\frac{\pi}{3}}^{m_{1}-m_{2}}\right){}^{*}-\left(t_{-\frac{\pi}{3}}^{-m_{1}}\right){}^{*}t_{-\frac{\pi}{3}}^{m_{2}-m_{1}}\right]\left[\left(t_{\frac{\pi}{3}}^{m_{2}-m_{3}}\right){}^{*}t_{\frac{\pi}{3}}^{-m_{3}}+t_{\frac{\pi}{3}}^{m_{3}-m_{2}}\left(t_{\frac{\pi}{3}}^{m_{3}}\right){}^{*}\right]\nonumber \\
+ & \left[t_{\frac{\pi}{3}}^{m_{1}}\left(t_{\frac{\pi}{3}}^{m_{1}-m_{2}}\right){}^{*}+\left(t_{\frac{\pi}{3}}^{-m_{1}}\right){}^{*}t_{\frac{\pi}{3}}^{m_{2}-m_{1}}\right]\left[\left(t_{-\frac{\pi}{3}}^{m_{2}-m_{3}}\right){}^{*}t_{-\frac{\pi}{3}}^{-m_{3}}+t_{-\frac{\pi}{3}}^{m_{3}-m_{2}}\left(t_{-\frac{\pi}{3}}^{m_{3}}\right){}^{*}\right]
\end{align}
It is easy to verify that it reduces to Eqs.~\eqref{eq:J_chi_b} and \eqref{eq:J_chi_generic}  in the monochromatic limit.

We next list the fourth-order corrections for $J_{1}$. For circular polarization,
$\delta J_{1}^{\left(4\right)}$ is

\begin{align}
\delta J_{1}^{\left(4\right)} & =\sum_{\boldsymbol{m}}8\mathcal{A}\left(\boldsymbol{m}\right)f_{\triangle}^{\left(CP\right)}\left(\boldsymbol{m}\right)-8\mathcal{L}\left(\boldsymbol{m}\right)\left[\cos\left(\frac{\pi m_{2}}{3}\right)+2\cos\left(\frac{2\pi m_{2}}{3}\right)\right]\nonumber \\
 & -16\mathcal{B}\left(\boldsymbol{m}\right)\left[\cos\left(\frac{\pi m_{2}}{3}\right)+2\cos\left(\frac{2\pi m_{2}}{3}\right)+2\right]-40\mathcal{G}\left(\boldsymbol{m}\right),\,\,\,\left(\text{CP}\right)\label{eq:dJ1-triang-CP}
\end{align}
with

\begin{align}
f_{\triangle}^{\left(CP\right)}\left(\boldsymbol{m}\right) & =2+\cos^{2}\left[\left(m_{1}+m_{3}\right)\frac{\pi}{2}\right]\left(\cos\left[\frac{1}{3}\pi\left(m_{1}-m_{3}\right)\right]+2\cos\left[\frac{2}{3}\pi\left(m_{1}-m_{3}\right)\right]\right)\nonumber \\
 & +\cos^{2}\left[\left(m_{1}+m_{2}+m_{3}\right)\frac{\pi}{2}\right]\left(\cos\left[\frac{1}{3}\pi\left(m_{1}-m_{2}+m_{3}\right)\right]+2\cos\left[\frac{2}{3}\pi\left(m_{1}-m_{2}+m_{3}\right)\right]\right).
\end{align}

The correction $\delta J_{1}^{\left(4\right)}$ to a bond along the
$\boldsymbol{\delta}_{3}$ direction coupled to linearly polarized
light is

\begin{align}
\delta J_{1}^{\left(4\right)} & =\sum_{\boldsymbol{m}}8\left\{ \cos^{2}\left[\left(m_{1}+m_{2}+m_{3}\right)\frac{\pi}{2}\right]+\cos^{2}\left[\left(m_{1}+m_{3}\right)\frac{\pi}{2}\right]\right\} \left(\mathcal{A}_{3,2,3,2}+\mathcal{A}_{2,3,3,2}+\mathcal{A}_{3,1,3,1}+\mathcal{A}_{1,3,3,1}\right)\nonumber \\
 & -8\cos^{2}\left[\left(m_{1}+m_{3}\right)\frac{\pi}{2}\right]\left(\mathcal{A}_{1,2,1,2}+\mathcal{A}_{2,1,1,2}\right)16\mathcal{A}_{2,2,2,2}-8\left(\mathcal{L}_{2,2,3,3}+\mathcal{L}_{3,3,2,2}-\mathcal{L}_{2,2,1,1}\right)\nonumber \\
 & -16\left[2\left(\mathcal{B}_{3,2}+\mathcal{B}_{2,3}\right)+2\mathcal{B}_{3,3}-\mathcal{B}_{1,2}\right]-8\left(2\mathcal{G}_{3,2}+2\mathcal{G}_{2,3}+2\mathcal{G}_{3,3}-\mathcal{G}_{1,2}\right),\,\,\,\left(\text{LP}\right)\label{eq:dJ1-triang-LP}
\end{align}
Notice that $\delta J_{1}^{\left(4\right)}\rightarrow-28t_{1}^{4}/U^{3}$
as $A_{0}\rightarrow0$, recovering the time-independent limit. The corrections for linearly polarized light
in other directions are found by permutations of the sub-indices.

In Fig.~\ref{fig:results2}, we show the modification of the exchange couplings as function of the fluence $A_0$ for two polarization protocols: by averaging over the entire Poincar\'e sphere (type I light) and by averaging over the equator of the sphere, consisting of an ensemble of linearly-polarized light (type II Glauber light). The main difference regards the ring-exchange term $J_{\square}$.  When the average is performed over the entire sphere, $J_{\square}$ becomes negative for $A_0=1.68$ before the maximum enhancement of $J_{2,3}$ is achieved. This poses a disadvantage as compared to the average over linear polarization when the goal is to destabilize the 120 phase and transition to a SL regime, but may lead to other phase transitions.

\begin{figure}

\includegraphics[width=1.0\columnwidth]{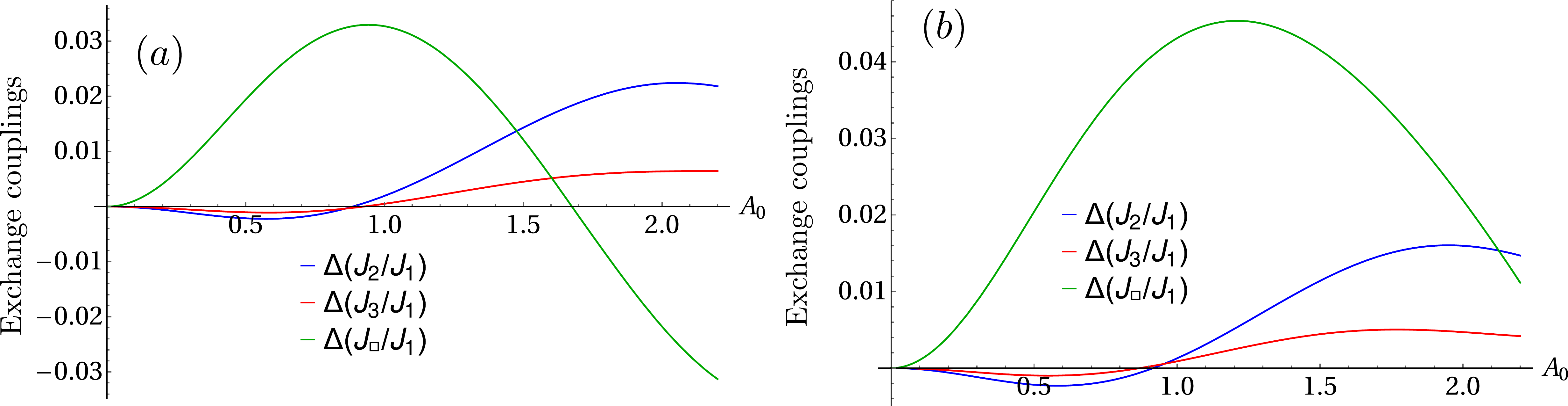}

\caption{Enhancement of magnetic couplings on the triangular lattice with varying fluence $A_{0}$ for two distinct protocols with $t_{1}/U=0.037$ and $\Omega/U=2/3$.  (a) Average over the entire Poincar\'{e} sphere (b) Average over linear polarization, the equator of the Poincar\'{e} sphere (see Fig.~1). The couplings $J_{2}$ and $J_{3}$ are initially decreased from their initial value, with $J_{2}$ becoming slightly negative (see Fig.~3 of the main text for the parametric plot of $J_{2}$ and $J_{3}$). In (a), the ring-exchange term $J_{\square}$ becomes negative before the enhancement of $J_{2,3}$ reaches the maximum value while in (b) the saturation of $J_{2,3}$ is before. \label{fig:results2}}
\end{figure}

One concern that arises from examining these corrections in Brillouin-Wigner theory is that we generically find terms in the denominator like $n U + m \Omega$, as found in Eq.~\ref{eq:L_m} for $n= 2$, where $m$ photons excite $n$ electrons across the Mott gap.  These naively suggest that there could be additional resonances for $\tilde{\Omega} =-n/m$ at every rational number.  However, these resonances do not appear due to the cancellation of contributions from different paths, in the Brillouin-Wigner theory.  To see that these \emph{always} vanish, it is necessary to go to the Schrieffer-Wolff transformation \cite{PolkovnikovPRL2016}, where it is evident that resonances only occur at $\tilde{\Omega} = -1/m$.

\subsubsection{Higher-order corrections}

We now comment about the effects corrections from higher orders in perturbation theory. Given that the odd powers of $t_{1}/U$ lead to vanishing contributions, the next finite order in perturbation theory is sixth order. By keeping the ratio $t_{1}/U<0.04$, as we must avoid heating, higher orders will contribute only small corrections to the fourth-order results. To justify the truncation of the perturbative expansion in the presence of the Floquet field, we may examine the relative contributions to $J_{1}$. Generically, there are two contributions: the second and the fourth-order ones, $J_{1}=J_{1}^{\left(2\right)}+J_{1}^{\left(4\right)}$. By computing the ratios $\left|J_{1}^{\left(2\right)}\right|/J_{1}$ and $\left|J_{1}^{\left(4\right)}\right|/J_{1}$ for the fluences considered in this work, 80\% or more of the total contribution to $J_{1}$ comes from the second-order term, $\left|J_{1}^{\left(2\right)}\right|/J_{1}\geq0.8$. For higher values of fluence, $J_{1}^{\left(2\right)}$ becomes small and can even pass through zero and go negative.  In this region, the sixth-order corrections must be incorporated, but otherwise are negligible.

\end{widetext}

\bibliographystyle{apsrev4-1}
\bibliography{FSLs_references.bib,biblio_Sahoo.bib}

\end{document}